\documentclass[aps,prl,reprint,groupedaddress,showpacs]{revtex4-1}
\usepackage{graphicx}
\usepackage{wasysym}

\newcommand{\ket}[1]{\ensuremath{\mid\!\! #1 \rangle}}
\begin{document}

% Use the \preprint command to place your local institutional report
% number in the upper righthand corner of the title page in preprint mode.
% Multiple \preprint commands are allowed.
% Use the 'preprintnumbers' class option to override journal defaults
% to display numbers if necessary
%\preprint{}

%Title of paper
\title{GHz Rabi flopping to Rydberg states in hot atomic vapor cells} 

% repeat the \author .. \affiliation  etc. as needed
% \email, \thanks, \homepage, \altaffiliation all apply to the current
% author. Explanatory text should go in the []'s, actual e-mail
% address or url should go in the {}'s for \email and \homepage.
% Please use the appropriate macro foreach each type of information

% \affiliation command applies to all authors since the last
% \affiliation command. The \affiliation command should follow the
% other information
% \affiliation can be followed by \email, \homepage, \thanks as well.
\author{B. Huber}
\thanks{First two authors contributed equally to this work.}
\affiliation{5. Physikalisches Institut, Universit\"{a}t Stuttgart,
Pfaffenwaldring 57, 70550 Stuttgart, Germany}
\author{T. Baluktsian}
\thanks{First two authors contributed equally to this work.}
\affiliation{5. Physikalisches Institut, Universit\"{a}t Stuttgart,
Pfaffenwaldring 57, 70550 Stuttgart, Germany}
\author{M. Schlagm\"uller}
\affiliation{5. Physikalisches Institut, Universit\"{a}t Stuttgart,
Pfaffenwaldring 57, 70550 Stuttgart, Germany}
\author{A. K\"olle}
\affiliation{5. Physikalisches Institut, Universit\"{a}t Stuttgart,
Pfaffenwaldring 57, 70550 Stuttgart, Germany}
\author{H. K\"ubler}
\affiliation{5. Physikalisches Institut, Universit\"{a}t Stuttgart,
Pfaffenwaldring 57, 70550 Stuttgart, Germany}
\author{R. L\"ow}
\affiliation{5. Physikalisches Institut, Universit\"{a}t Stuttgart,
Pfaffenwaldring 57, 70550 Stuttgart, Germany}
\author{T. Pfau}
%\email[]{t.pfau@physik.uni-stuttgart.de}
%\homepage[]{www.pi5.uni-stuttgart.de}
\affiliation{5. Physikalisches Institut, Universit\"{a}t Stuttgart,
Pfaffenwaldring 57, 70550 Stuttgart, Germany}

%Collaboration name if desired (requires use of superscriptaddress
%option in \documentclass). \noaffiliation is required (may also be
%used with the \author command).
%\collaboration can be followed by \email, \homepage, \thanks as well.
%\collaboration{}
%\noaffiliation

\date{\today}

\begin{abstract}
We report on the observation of Rabi oscillations to a Rydberg state on a timescale below one nanosecond in thermal rubidium vapor. We use a bandwidth-limited pulsed excitation and observe up to 6 full Rabi cycles within a pulse duration of $\sim$ 4 ns. We find good agreement between the experiment and numerical simulations based on a surprisingly simple model. This result shows that fully coherent dynamics with Rydberg states can be achieved even in thermal atomic vapor thus suggesting small vapor cells as a platform for room temperature quantum devices. Furthermore the result implies that previous coherent dynamics in single atom Rydberg gates can be accelerated by three orders of magnitude.
\end{abstract}

% insert suggested PACS numbers in braces on next line
% insert suggested keywords - APS authors don't need to do this
%\keywords{}

%\maketitle must follow title, authors, abstract, \pacs, and \keywords
\maketitle

The properties of Rydberg atoms are dominated by very loosely bound electrons moving on large orbits around a charged nucleus. They can easily be a few thousand times larger in size than ground state atoms. Due to the weak bond, even small electric fields can induce large dipole moments and therefore give rise to strong interactions between Rydberg atoms. This in turn leads to strong optical nonlinearities and collective quantum behavior in a frozen  ensemble of atoms as e.g.\ the excitation blockade into Rydberg states \cite{Tong2004,Singer2004,Vogt2006}. These strong interactions are the basis for various applications of Rydberg atoms in quantum information processing \cite{Jaksch2000,Lukin2001,Saffman2010} and sensing \cite{Honer2011}. Most recently two seminal experiments demonstrated with the help of the Rydberg blockade first gate operations between two individually trapped ultracold atoms \cite{Isenhower2010,Wilk2010}. In these experiments the bandwidth of the coherent optical excitation rate, and by this the time of the gate operation, has been limited to the MHz scale. The limit is mainly due to the small spatial overlap between low lying electronic states and the extended Rydberg states and due to the usage of medium power cw laser systems. It is of fundamental importance to increase the speed of the coherent excitation to allow for more complex algorithms during the limited time set by decoherence processes. One way to overcome this technical limitation is the use of coherent i.e.\ bandwidth-limited pulsed laser systems for the excitation. We show that Rabi cycles to Rydberg states below one nanosecond are feasible, resulting in a speedup by three orders of magnitude. 

For such short evolution times even atoms at room temperature can be used to observe Rabi flopping to Rydberg states as they can be considered frozen on this timescale. This opens up new horizons for room temperature quantum devices based on the ensemble Rydberg blockade which have been described in various proposals \cite{Saffman2010,LowR.Pfau2009,Saffman2002,Honer2011}. First steps towards the fabrication of microscopic vapor cells have already been accomplished \cite{Baluktsian2010,Sarkisyan2001}. Standard technologies in the production process of these cells will allow for full integration of larger cell arrays \cite{Yang2007} and applications e.g.\ for linear quantum computing \cite{Politi2008} are in reach. 
From previous work on electromagnetically induced transparency in thermal vapor involving Rydberg states, a coherence time limit of a few 10 nanoseconds can be inferred \cite{Mohapatra2007,Kubler2010} which is due to transient broadening and wall interactions. 

In this letter we report on the observation of Rabi oscillations to a Rydberg state in hot atomic vapor. We accomplish this by driving the dynamics fast compared to any decoherence processes. We achieve Rabi cycle times of $\sim\! 500\,\mathrm{ps}$ resulting in a coherent phase evolution of up to $12\pi$ which is limited only by the length of the pulse used in the experiment.
In the following we first briefly describe the experimental setup and measuring scheme we used to obtain the oscillation data which is presented subsequently. The behavior of the oscillations for different driving intensities and laser detunings is shown. We give qualitative arguments for the behavior and compare with numerical simulations based on a very simple model.

\begin{figure}
\centering
\includegraphics[width=\columnwidth]{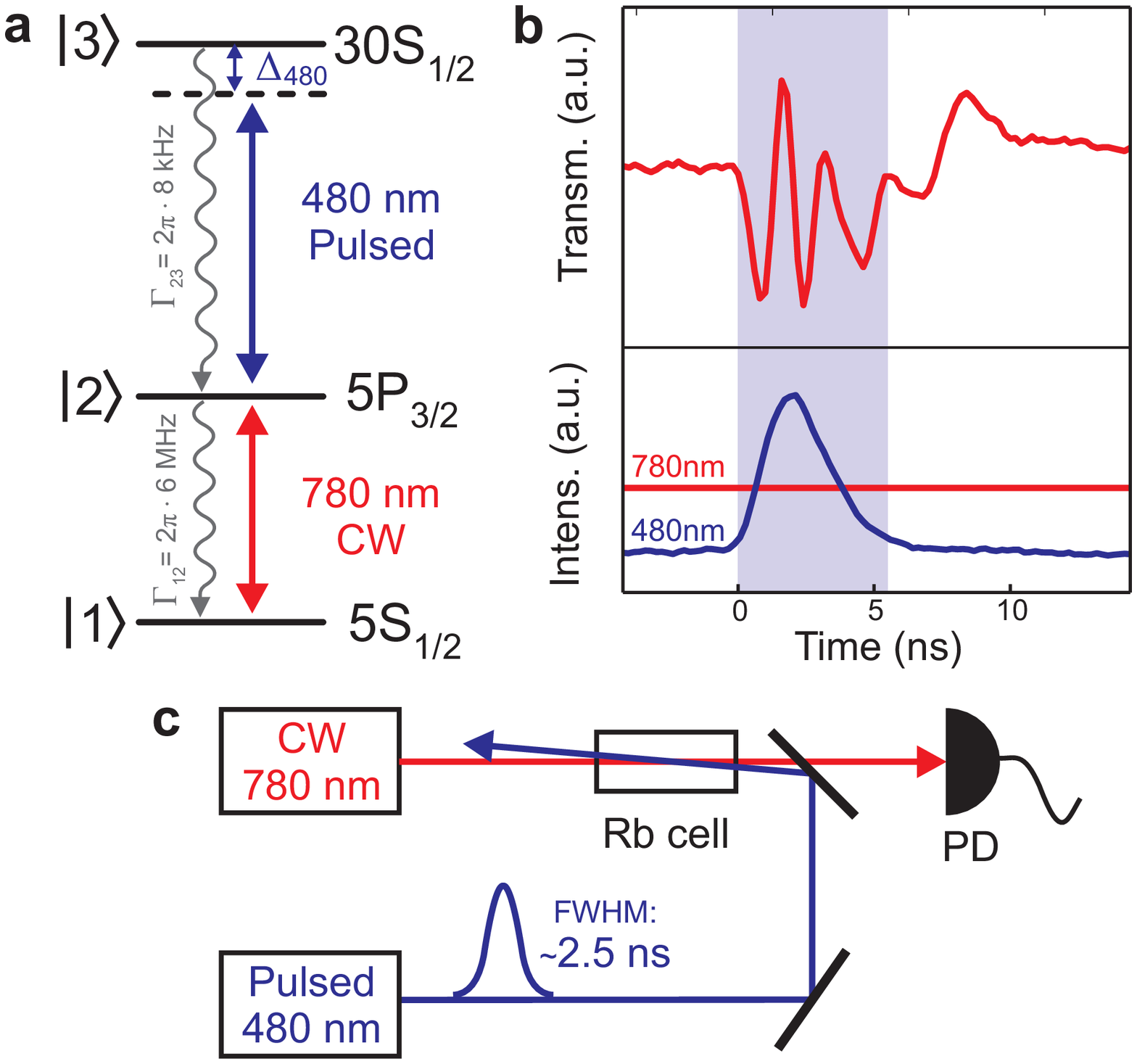}
\caption{\textbf{Schema of the experiment.} 
 (a) The Rydberg state is addressed with a two-photon-transition via an intermediate state. The laser for the upper transition is pulsed. (b) Transmission signal of the $780\,\mathrm{nm}$ laser (upper half) and corresponding laser intensities (lower half) as a function of time. (c) Simplified schematic of the optical setup.
\label{fig:schema}}
\end{figure}

For the following experiments we use a $5\,\mathrm{mm}$ glass cell containing rubidium vapor at $T \sim 130\,\mathrm{^\circ C}$ with the two stable isotopes occurring at natural abundance ($^{85}\mathrm{Rb:}\, 72.2\%$ and $^{87}\mathrm{Rb:}\, 27.8\%$). The atomic density was determined by absorption spectroscopy to be ${7.4 \cdot 10^{12}\,\mathrm{cm^{-3}}}$. We address the Rydberg state in $^{85}\mathrm{Rb}$ using a two-photon-transition via an intermediate state. The excitation scheme is $5S_{1/2}\rightarrow 5P_{3/2}\rightarrow 30S_{1/2}$, denoted by \ket{1}, \ket{2} and \ket{3} in the following (Fig.~\ref{fig:schema} (a)). The ground state transition is driven by a cw laser at $\sim\! 780\,\mathrm{nm}$. This laser is locked to the center of the Doppler valley of the $^{85}\mathrm{Rb}, 5S_{1/2}, F=2\rightarrow 5P_{3/2}$ transition in order to ensure a large spectral distance from any other ground state transition (from either isotope). For the upper transition to the Rydberg state a bandwidth-limited laser pulse at $\sim\! 480\,\mathrm{nm}$ is employed. The pulse is created with a seeded dye amplifier in four-pass configuration similar to the one described in \cite{Schwettmann2007}. The temporal shape of the pulse is roughly Gaussian with a full width of $\sim\!2.5\,\mathrm{ns}$ at half maximum (Fig.~\ref{fig:schema} (b), lower half).
%High enough Rabi frewuencies
The corresponding Rabi frequencies are $\Omega_{780}\sim\! 2\pi\cdot 220\,\mathrm{MHz}$ for the cw light and $\Omega_{480, \mathrm{peak}}\sim\! 2\pi\cdot 2.2\,\mathrm{GHz}$ for the maximum peak Rabi frequency of the pulse \cite{Note1}. Note that all Rabi frequencies are much larger than the decay rate of any state involved such that the dynamics is not dominated by decay on the relevant timescales. The two laser beams are overlapped in the vapor cell in almost counter-propagating ($\theta = 171.5\mathrm{^\circ}$) configuration (Fig.~\ref{fig:schema} (c)). We observe the transmission of the $780\,\mathrm{nm}$ laser with a fast photodetector thus probing the coherence between the lower two levels. In order to avoid averaging over different Rabi frequencies we ensure that only light from a homogeneously illuminated region of the cell is detected. We achieve this by selecting an almost homogeneous intensity distribution in the blue beam profile with a circular pinhole and image it into the cell via 1:1 imaging. We then cut out a part of the illuminated volume by use of another 1:1 imaging onto a second pinhole which resides directly in front of the photodetector. This way we detect a homogeneously illuminated region also with respect to the $780\,\mathrm{nm}$ laser. This region contains about $3\cdot 10^8$ atoms.

Both lasers are linearly polarized in the same direction. Note that due to the large bandwidth of the excitation dynamics, the hyperfine structure in the $5P_{3/2}$ and the $30S_{1/2}$ state is not resolved while it is resolved in the $5S_{1/2}$ ground state. This means that the typical inhomogeneous distribution of matrix elements in narrow band experiments \cite{Reetz-Lamour2008} (e.g. in the MHz range as in cold atom experiments) is not present here and the medium reduces in good approximation to a thermal ensemble of driven three level atoms despite the initial mixture of hyperfine ground states. 

A typical transmission curve we observe (averaged over 5 curves) is plotted in the upper half of fig.~\ref{fig:schema} (b).  Before the pulse, the atoms represent an effective 2-level-system (not taking into account the second hyperfine ground state) and are in a steady state together with the strongly driving $780\,\mathrm{nm}$ light. During the time of the pulse (depicted by the shaded area), oscillations in the transmission appear. For a resonant pulse, these oscillations mainly correspond to Rabi flopping between the intermediate and the Rydberg state. Depending on the phase of the oscillation at the end of the pulse, a certain fraction of the occupation remains in the Rydberg state. Further oscillatory behavior in the time after the pulse originates from Rabi oscillations between the lower two states as the remaining 2-level-system is not in equilibrium immediately after the pulse. Those oscillations decay quickly and the system returns into a steady state.

\begin{figure}
\centering
\includegraphics[width=\columnwidth]{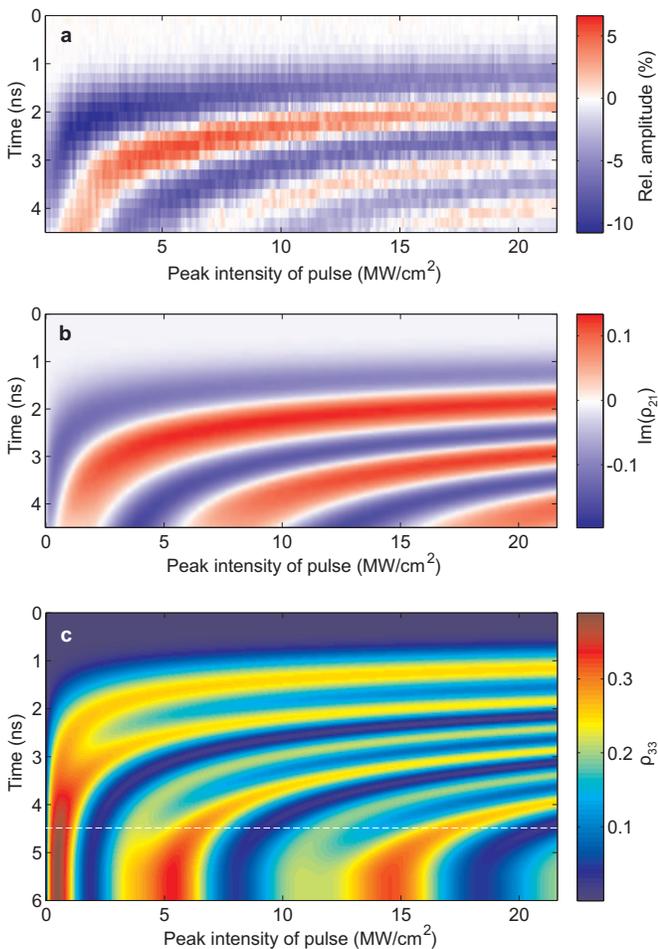}
\caption{\textbf{Rabi oscillations for different pulse intensities.} 
(a) Oscillations in the transmission of the $780\,\mathrm{nm}$ laser during the pulse. Both lasers are resonant to their respective transitions. The beginning of the pulse is at $t=0\,\mathrm{ns}$.
(b) A corresponding simulation provides results for $\mathrm{Im}(\rho_{21})$ which is connected to the absorption. The behavior of the experimental data is reproduced in excellent agreement. 
(c) Simulated Rydberg occupation $\rho_{33}$. Above the dashed line the dynamics is dominated by the pulsed laser. All relevant parameters are described in the text.
\label{fig:powerscans}}
\end{figure}

We studied the behavior of the system for different pulsed laser intensities (Fig.~\ref{fig:powerscans} (a)). With both lasers on resonance, we recorded transmission curves for pulsed laser peak intensities from $I_{480, \mathrm{peak}} = 0\ldots 21\,\mathrm{MW/cm^2}$. With the corresponding dipole matrix element (taken from \cite{Saffman2010}) this translates to Rabi frequencies of $\Omega_{480, \mathrm{peak}} \sim\! {0\ldots 2\pi\cdot 2.3\,\mathrm{GHz}}$. Each time trace in the figure is an average of 5 transmission curves. For a 3-level-system in ladder-configuration with both transitions resonantly driven, one expects the eigenfrequencies to behave like $\Omega_{\mathrm{eff}, \mathrm{res}} \propto \sqrt{\Omega_{780}^2 + \Omega_{480}^2}.$ Indeed, points of constant phase exhibit this square root-like behavior as a function of pulse intensity. 

We have done numerical simulations of a 3-level-system with a density matrix approach using the Liouville-von Neumann equation \cite{Fleischhauer2005} 
$$\frac{\partial \hat{\rho}}{\partial t} = -\frac{i}{\hbar} \left[ \hat{H}, \hat{\rho}\right] + \hat{L}(\hat{\rho})$$
with the Hamiltonian in the rotating wave approximation
$$\hat{H} = \hbar
\left(
\begin{array}{ccc}
0													&\frac{1}{2}\Omega_{780} 			&0\\
\frac{1}{2}\Omega_{780}^*	&-\Delta_{780}								&\frac{1}{2}\Omega_{480}(t)\\
0													&\frac{1}{2}\Omega_{480}^*(t)	&-\Delta_{780}-\Delta_{480}
\end{array}
\right)
$$
and the Liouville operator
%$$\hat{L}(\hat{\rho}) = 
%\left(
%\begin{array}{ccc}
%\Gamma_{12} \rho_{22}							&-\frac{1}{2}\Gamma_{12} \rho_{12}							&-\frac{1}{2}\Gamma_{23} \rho_{13}\\
%-\frac{1}{2}\Gamma_{12} \rho_{12}	&-\Gamma_{12} \rho_{22}+\Gamma_{23} \rho_{33}		&-\frac{1}{2}(\Gamma_{12}+\Gamma_{23})\rho_{23}\\
%-\frac{1}{2}\Gamma_{23} \rho_{31}	&-\frac{1}{2}(\Gamma_{12}+\Gamma_{23})\rho_{32}	&-\Gamma_{23} \rho_{33}
%\end{array}
%\right)
%$$
\begin{eqnarray*}
\hat{L}(\hat{\rho}) = 
\Gamma_{12}
\left(
\begin{array}{ccc}
 \rho_{22}						&-\frac{1}{2} \rho_{12}	&0\\
-\frac{1}{2}\rho_{21}	&-\rho_{22}							&-\frac{1}{2}\rho_{23}\\
0											&-\frac{1}{2}\rho_{32}	&0
\end{array}
\right)\\
+
\Gamma_{23}
\left(
\begin{array}{ccc}
0											&0											&-\frac{1}{2}\rho_{13}\\
0											& \rho_{33}							&-\frac{1}{2}\rho_{23}\\
-\frac{1}{2}\rho_{31}	&-\frac{1}{2}\rho_{32}	&-\rho_{33}
\end{array}
\right)
\end{eqnarray*}
that includes the decay rates $\Gamma_{12} = 2\pi \cdot 6\,\mathrm{MHz}$ \cite{Volz1996} and $\Gamma_{23} = 2\pi \cdot 8\,\mathrm{kHz}$ \cite{Branden2010}. The initial conditions are given by the steady state solution ($\partial \hat{\rho} / \partial t = 0$) in the absence of the pulsed laser ($\Omega_{480} = 0$). We account for the thermal distribution of the atoms by solving this equation separately for each velocity class with the respective Doppler detunings and weighting the individual results accordingly. The quantity connected to the absorption is the imaginary part of the coherence of the $\ket{1}\rightarrow\ket{2}$ transition $\mathrm{Im}(\rho_{21})$ (Fig.~\ref{fig:powerscans} (b)). For small relative variations proportionality between the transmission and $\mathrm{Im}(\rho_{21})$ can be assumed. The behavior of $\mathrm{Im}(\rho_{21})$ is found to be in excellent agreement with the experiment.

While our experimental approach does not allow for direct observation of the Rydberg population $\rho_{33}$, this quantity is accessible in the simulations (Fig.~\ref{fig:powerscans} (c)). It can be seen that the population oscillates at twice the frequency that can be observed in the transmission. This is expected as the atomic state acquires a phase of $\pi$ for each round trip $\ket{2} \rightarrow \ket{3} \rightarrow \ket{2}$. In contrast to $\mathrm{Im}(\rho_{21})$, $\rho_{33}$ is not sensitive to phases. Therefore, the evolution of $\mathrm{Im}(\rho_{21})$ recurs after two round trips, whereas the evolution of $\rho_{33}$ recurs after one. The data shows that during the time of the pulse ($\sim\! 4\,\mathrm{ns}$) we achieved to drive Rabi cycles with a phase of up to $12 \pi$.

After the pulse, the Rydberg population stays nearly constant due to the long lifetime of the Rydberg state. The lifetime is limited by Rydberg - ground state collisions and transient broadening which are not included in the simulations but expected to be on the order of few MHz \cite{Thompson1987}. By adjusting the pulse intensity, the phase of the oscillation at the end of the pulse can be chosen conveniently such that a maximum Rydberg population remains after the pulse. For our experimental parameters, where the experiment was not optimized for high Rydberg population, we can infer that $\rho_{33} \sim\! 35\%$. This is, besides thermal velocity distribution, due to the fact that we start with a steady state, where the intermediate state population is $\sim\! 50\%$. 
With a sequence, however, where also the lower transition is addressed by a pulse, the limit of $50\%$ does not apply anymore. Impromptu simulations of simultaneous pulses for both transitions with Rabi frequencies in the GHz range show that for pulse lengths of $\lesssim 1\,\mathrm{ns}$ Rydberg occupations of over 90\% can easily be achieved.

% and the fraction of atoms, that can be pumped to the Rydberg state is mainly determined by the bandwidth of the pulses. Proof-of-principle calculations of short enough ($<1\,\mathrm{ns}$) simultaneous pulses for both transitions show that Rydberg occupations of over 90\% can be easily achieved.

% Proof-of-principle calculations of simultaneous pulses for both transitions show that Rydberg occupations of over 90\% can be easily achieved ($\Omega_{480} = \Omega_{780} = 2\pi \cdot 2\,\mathrm{GHz}$; width: $0.5\,\mathrm{ns}$).

%This limit of $50\%$, however, can be overcome by using a sequence where also the lower transition is addressed by a pulse. Further simulations indicate that Rydberg occupations of well over 90\% can be achieved by using a sequence where also the lower transition is addressed by a pulse.

\begin{figure}
\centering
\includegraphics[width=\columnwidth]{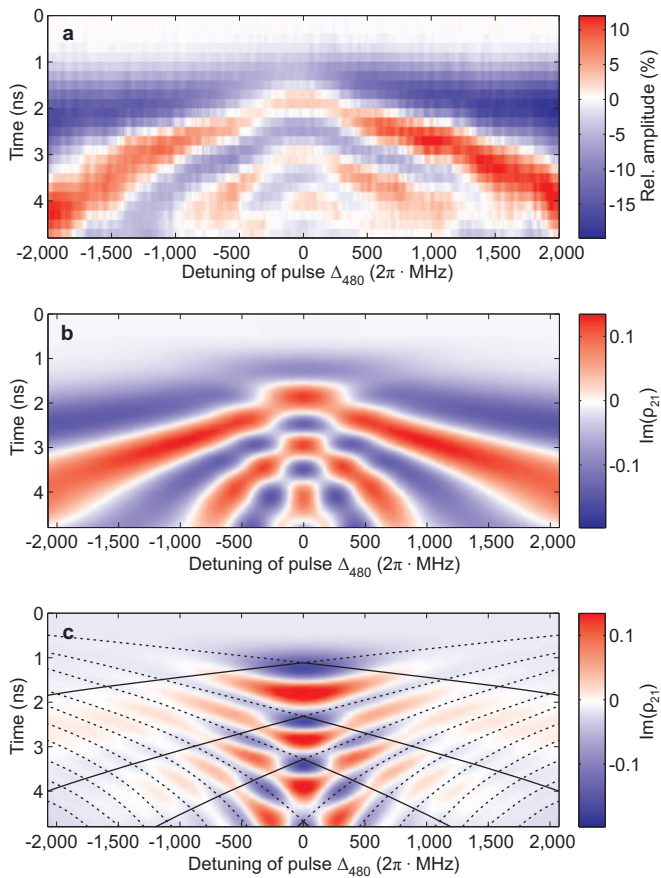}
\caption{\textbf{Rabi oscillations for different detuning of the pulse.} 
(a) Oscillations in the transmission of the $780\,\mathrm{nm}$ laser during the pulse. This laser is resonant to the $\ket{1}\rightarrow\ket{2}$ transition. The beginning of the pulse is at $t=0\,\mathrm{ns}$. %(5\%~threshold)
(b)~Simulation of $\mathrm{Im}(\rho_{21})$ which is connected to the absorption. The behavior agrees very well with the experimental data. (c)~Simulated behavior of $\mathrm{Im}(\rho_{21})$ for the velocity class $v = 0$ only. The black lines (both dashed and solid) indicate points of constant phase for the two oscillatory modes that occur.
\label{fig:freqscans}}
\end{figure}

Furthermore we investigated the dependence of the system on the detuning of the pulsed laser. For these measurements we used a peak pulse intensity of $21\,\mathrm{MW/cm^2}$.
The oscillations observed in the transmitted beam (Fig.~\ref{fig:freqscans}(a)) are reproduced in very good agreement by simulations (Fig.~\ref{fig:freqscans} (b)). 
It can be seen that the oscillations become slower with increasing detuning. This behavior may seem counter-intuitive at first glance as Rabi oscillations in a 2-level-system scale like $\Omega_{\mathrm{eff}} \propto \sqrt{\Omega^2+\Delta^2}$, thus get faster with detuning. In our system, however, the laser pulse induces a dynamic Autler-Townes splitting between the states \ket{2} and \ket{3}.
Two oscillatory modes occur, one arising from each of the split lines \cite{Rzazewski1984}.
For an atom at rest (Fig.~\ref{fig:freqscans} (c)) the two modes (indicated by the black lines of positive and negative slope, respectively) are symmetric and the highest oscillation amplitude can be found for zero pulse detuning. For increasing pulse detuning, one of the split lines approaches the resonance. The corresponding mode therefore becomes slower (solid part of the lines) while the other mode becomes faster (dashed part of the lines).
For atoms with a finite velocity, however, both lasers appear Doppler shifted such that resonance to one of the split lines can actually be reached for a certain pulse detuning. Hence, the slower mode becomes more intense as the pulse detuning approaches this resonance. At the same time the faster mode becomes more faint. 
Additionally, the visibility of the faster mode is reduced further by the fact that the coupling is varying in time due to the envelope of the pulse.
Averaging over all velocity classes then yields the observed behavior (Fig.~\ref{fig:freqscans} (a), (b)).

In summary, we have found that the observed GHz Rabi oscillations to a Rydberg state in thermal vapor show a perfect match with a very simple theoretical model based on 3 levels only. This is surprising at first sight as rubidium has a complex hyperfine structure. However, the large bandwidth of the fast excitation greatly simplifies the situation as compared to slow dynamics which suffers from a state dependent inhomogeneous distribution of coupling strengths to the light field. This reduces the necessity for initialization of ensembles in quantum devices. It also simplifies the physics of the Rydberg blockade. In future experiments we therefore intend to use the observation of coherent dynamics to demonstrate the Rydberg blockade and the characteristic $\sqrt{N}$-scaling of the collective Rabi frequency in thermal vapor as has been done in ultracold ensembles previously \cite{Heidemann2007,Urban2009,Pritchard2010}.
We envisage the application of coherent dynamics in Rydberg blockaded thermal ensembles in the realization of single photon emitters \cite{Saffman2002} and absorbers \cite{Honer2011}.

% Specify following sections are appendices. Use \appendix* if there
% only one appendix.
%\appendix
%\section{}

% If you have acknowledgments, this puts in the proper section head.
\begin{acknowledgments}
We acknowledge fruitful discussions with H.P.~B\"uchler, H.~Giessen, K.~Rz\c a\.zewski and valuable advice from J.~Shaffer. The work is supported by the ERC, BMBF, MALICIA and contract research 'Internationale Spitzenforschung' of the Baden-W\"urttemberg Stiftung. B.H. acknowledges support from Studienstiftung des deutschen Volkes. T.B. acknowledges support from the Carl Zeiss Foundation.
\end{acknowledgments}

% Create the reference section using BibTeX:
%

\end{document}